\documentclass[a4paper,12pt]{spieman}  
\usepackage{amsmath,amsfonts,amssymb}
\usepackage{aas_macros}
\usepackage{graphicx}
\usepackage[dvipsnames]{xcolor}
\usepackage{setspace}
\usepackage{tocloft}

\title{Black hole accretion, star formation, and chemical evolution with PRIMA/FIRESS spectroscopy: toward the cosmic noon and beyond}

\author[a,b,*]{Juan Antonio Fernández-Ontiveros}
\author[b]{Luigi Spinoglio}
\author[c]{Tohru Nagao}
\affil[a]{Centro de Estudios de F\'isica del Cosmos de Arag\'on (CEFCA), Plaza San Juan 1, E--44001 Teruel, Spain}
\affil[b]{Istituto di Astrofisica e Planetologia Spaziali (INAF--IAPS), Via Fosso del Cavaliere 100, I--00133 Roma, Italy}
\affil[c]{Research Center for Space and Cosmic Evolution, Ehime University, 2-5 Bunkyo-cho, Matsuyama, Ehime 790-8577, Japan}

\cftpagenumbersoff{figure}
\cftpagenumbersoff{table}

\DeclareRobustCommand{\ion}[2]{\textup{#1\,\textsc{\lowercase{#2}}}}
\newcommand*\farcs{\ensuremath{\overset{\prime\prime}{.}}}

\begin{document} 
\maketitle

\begin{abstract}
The PRobe far-Infrared Mission for Astrophysics (PRIMA) will provide the first complete three-dimensional view of the hidden side of star formation and black hole accretion that dominate energy production in galaxies during the cosmic noon. Nearly 90\% of UV/optical photons from young stars in galaxies and Active Galactic Nuclei (AGN) at cosmic noon are absorbed by dust and reradiated in the mid- to far-infrared range. The latter hosts a unique suite of diagnostic lines for tracing black hole accretion and star formation, and probing physical and chemical conditions in galaxies. PRIMA will conduct blind spectroscopic surveys out to $z \sim 3$--$4$, enabling the first statistically unbiased determination of star formation and black hole accretion co-evolution with cosmic time.

In this work, we simulate a 200 arcmin$^2$ blind spectroscopic survey with FIRESS to measure star formation and black hole accretion rates for hundreds of galaxies during cosmic noon. Follow-up observations of selected targets from this or other surveys will determine relative N/O abundances and N/O-independent metallicities using multiple mid-IR emission lines. Using FIRESS's high-resolution mode, it will be possible to study feedback mechanisms by detecting P-Cygni profiles and blueshifted absorption wings in molecular lines such as OH, and through emission line wings of highly ionized gas.

Beyond cosmic noon, PRIMA will pioneer rest-frame mid-IR spectroscopy of galaxies in the reionization epoch. Bright high-ionization lines in the mid-IR will provide a transformative view of the primary ionization continuum of bright Lyman-$\alpha$ emitters and low-metallicity AGN in the early Universe, revealing conditions that drove the first generations of galaxies and black holes. By filling a critical observational gap between optical and radio wavelengths left by existing and planned facilities, PRIMA will deliver a comprehensive and unobscured view of the processes governing galaxy evolution, from the earliest epochs to the peak of star formation activity.
\end{abstract}


{\noindent \footnotesize\textbf{*}\linkable{j.a.fernandez.ontiveros@gmail.com}, \linkable{jafernandez@cefca.es}}


\section{Introduction}\label{sec_intro}
Deep infrared (IR) spectroscopic surveys, unaffected by dust obscuration, are essential to study galaxy evolution by tracing both star formation and supermassive black hole accretion in active galactic nuclei (AGN) across space and time. Both processes exhibit parallel evolution with redshift, peaking during cosmic noon ($z \sim 1$--$3$) before declining steeply to the present epoch\cite{madau_dickinson2014}. Furthermore, the interaction between AGN and star formation during this period is often invoked to explain the main properties of present-day galaxies, including: the deviation of the galaxy stellar mass function from that of dark matter halos\cite{Silk2012,Wechsler2018,Behroozi2019}, the transition of star-forming galaxies from the blue cloud to the red sequence\cite{Schawinski2014,Heckman2014}, and several scaling relations in galaxies, such as those observed between the mass of central supermassive black holes (SMBHs) and the bulge stellar velocity dispersion\cite{Magorrian1998,Ferrarese2000}, the luminosity and the rotation velocity or the bulge stellar velocity dispersion\cite{Tully1977,Faber1976}, and the stellar mass-metallicity relation\cite{Lequeux1979,Tremonti2004,Gallazzi2005}. Therefore, understanding galaxy evolution requires measuring both SFR and black hole accretion rate (BHAR) across cosmic time.

However, during the peak epoch of star formation, most galaxies experience heavy absorption of UV and optical light by dust, which re-emits approximately 90\% of the energy released by young stars in the IR \cite{madau_dickinson2014}. Observations of the UV rest-frame continuum associated with young massive stars (1400--1700\,\AA), in spite of being one of the most popular methods to measure the star-formation rate (SFR) density\cite{Lilly1995,Madau1996,Cucciati2012,Bouwens2015}, is highly unreliable in obscured environments such as those typical of galaxies during the cosmic noon. Therefore, rest-frame ultraviolet (UV) and optical observations cannot access these crucial dust-obscured regions where massive stars form.

Another essential aspect to understand galaxy evolution is the production of heavy elements, which is intrinsically linked to the processes governing the assembly of stellar mass in galaxies. The gas metallicity is particularly sensitive to the past star-formation history, the gas inflows from the surrounding environment, and the gas outflows due to energetic events such as starburst and AGN activity\cite{Maiolino2019}. So far, rest-frame optical diagnostics\cite{Nagao2006} have been used to measure the ISM metallicity of many star-forming galaxies up to very high-z, and consequently the redshift evolution of the mass-metallicity relation has been reported\cite{Maiolino2008}. Nevertheless, rest-frame optical diagnostics are effective only for optically thin galaxies. 
This limitation is evidenced by massive ($>$ $10^{10.5} M_{\odot}$) dusty galaxies (ultra-luminous infrared galaxies; ULIRGs) in the local Universe, which exhibit optical metallicities significantly lower than predicted by both the mass-metallicity relation\cite{Rupke2008,Caputi2008} and their dust-to-gas ratios\cite{Santini2010,Rowlands2014}. Thus, optical diagnostics may only probe the chemically younger, optically thin regions in galaxy outskirts, while IR diagnostics are required to access the metal-enriched, dust-obscured ISM\cite{Chartab2022,Perez-Diaz2024}. On the other hand, IR metallicities measured in galaxies at high redshift are primarily based on the ratios of nitrogen and oxygen far-IR fine-structure lines\cite{Lamarche2018,Rigopoulou2018}, with the N3O3 index --\,derived from the [\ion{O}{iii}]$_{\rm 52,88\mu m}$ and [\ion{N}{iii}]$_{\rm 57 \mu m}$ lines (hereafter [\ion{O}{iii}]$_{52,88}$ and [\ion{N}{iii}]$_{57}$, respectively)\,-- being the most reliable tracer\cite{Nagao2011,Pereira-Santaella2017,Peng2021,Spinoglio2022a}. However, deviations from the local O/H--N/O calibration\cite{Peng2021,Spinoglio2022a,Perez-Diaz2024}, including unexpectedly high N/O abundances like those recently discovered in very high-z galaxies\cite{Topping2024,Topping2025}, suggest a more chemically complex scenario that requires a more comprehensive exploitation of N/O-independent, IR-based metallicity diagnostics.

A complete understanding of galaxy evolution through cosmic noon, including the dominant obscured galaxy population, requires a sensitive, cryogenically-cooled space telescope with a wide infrared wavelength coverage\cite{jafo2017}. Such a facility will address the following key questions:
\begin{enumerate}
   \it
   \item What physical processes led to the peak of star formation and black hole growth at cosmic noon ($z \sim 2$), and the subsequent rapid quenching of star formation at $z < 1$?
   
   \item How do star formation, black hole accretion, and feedback from AGN and supernovae shape galaxy evolution, determining the properties of local Universe galaxies?
   
   \item How has metal enrichment evolved in galaxies across cosmic time?
\end{enumerate}

This work explores the potential role of PRIMA (The PRobe far-Infrared Mission for Astrophysics; Glenn et al., this volume) in our understanding of galaxy evolution. PRIMA is a proposed 1.8\,m cryogenically-cooled IR spectrocopic telescope selected for the Probe Explorers Phase A study of NASA's Explorers Program. The onboard Far-IR Enhanced Survey Spectrometer (FIRESS; Bradford et al., this volume) is ideally suited for this task, offering a broad wavelength coverage ($24$--$235\,\mu\text{m}$) and an unprecedented sensitivity ($\sim 1.5\,\text{dex}$) and mapping speed (1\,000--100\,000 improvement) over previous far-IR observatories such as Herschel\cite{Pilbratt2010} and SOFIA\cite{Temi2014}. The main advantages of the FIRESS spectrometer with respect to the Herschel/PACS\cite{Poglitsch2010} and SOFIA/FIFI-LS\cite{Fisher2018} spectrometers include both the multiplexing capability and the detector sensitivity.

\section{A blind spectroscopic survey}

Mid- to far-IR spectroscopy from space with PRIMA, using the FIRESS instrument in the low resolution mode (R$\sim$100, always $>85$), can answer the previous questions. PRIMA will quantify the contribution of AGN to the overall energy budget of heavily dust-obscured galaxies, measuring their accretion rate through the [\ion{O}{iv}]$_{\rm 25.9 \mu m}$ line (hereafter [\ion{O}{iv}]$_{25.9}$). The high-ionization potential of O$^{3+}$ ($54.9\, \rm{eV}$) guarantees a negligible contamination by star forming activity, and thus is the brightest IR line tracing genuine AGN activity, as demonstrated by several studies of AGN in the local Universe\cite{Melendez2008,Rigby2009,Dicken2014,Mordini2021,Spinoglio2022b}. Similarly, the contribution from hot, young stars, will be traced by the SFR obtained from the emission bands of polycyclic aromatic hydrocarbons (PAH) and the brightest nebular fine structure lines of, e.g. [\ion{Ne}{ii}]$_{\rm 12.8 \mu m}$ ([\ion{Ne}{ii}]$_{12.8}$), [\ion{Ne}{iii}]$_{\rm 15.6 \mu m}$ ([\ion{Ne}{iii}]$_{15.6}$), [\ion{S}{iii}]$_{\rm 18.7,33.5 \mu m}$ ([\ion{S}{iii}]$_{18.7,33.5}$), [\ion{S}{iv}]$_{\rm 10.5 \mu m}$ ([\ion{S}{iv}]$_{10.5}$), and [\ion{O}{iii}]$_{52,88}$. In particular, PAH molecules contribute to $5$--$20$\% of the emission in metal-rich dusty galaxies\cite{Smith2007}, and thus SFR calibrations based on these features become the most sensitive tracers in those environments\cite{Roussel2001,Kennicutt2012,Xie2019,Mordini2021}. They are also selective tracers, since they tend to be destroyed in the vicinity of AGN\cite{Siebenmorgen2004,Smith2007}. On the other hand, low-metallicity galaxies are particularly bright [\ion{Ne}{iii}]$_{15.6}$ emitters\cite{Cormier2015,Cormier2019}, and thus robust SFR calibrations have been based on this line for local samples of metal-poor galaxies\cite{Ho2007,Xie2019,Mordini2021}. Overall, the potential of IR diagnostics involving these lines and PAH bands was theoretically predicted and demonstrated in the local Universe\cite{Spinoglio1992,Genzel1998,Sturm2002,Armus2007,Veilleux2009,jafo2016} with ISO\cite{Kessler1996,deGraauw1996,Clegg1996}, Spitzer\cite{Werner2004,Houck2004}, and Herschel. No other currently planned telescope will perform such detailed spectroscopic investigations. PRIMA will fill the crucial spectral gap between JWST and ALMA in the $20$--$250\, \rm{\mu m}$ range.

Our approach is to conduct a blind spectroscopic survey of galaxies and AGN at cosmic noon. Spectroscopic surveys offer several key advantages: they efficiently detect galaxies with extreme emission lines indicative of intense star formation and nuclear activity, which are common at high-z and are crucial for understanding the ISM of early Universe galaxies\cite{Bunker2024,Meyer2024,Boyett2022,Chen2025,Llerena2024,Bugiani2025}. Many extreme line emitters, despite their faint continuum emission, are often missed in broad-band photometric surveys\cite{Lumbreras-Calle2022}. Furthermore, accurate spectroscopic redshifts enable environmental studies critical for understanding stellar and SMBH growth and activity quenching in dense environments like galaxy clusters and groups\cite{Shah2024,Galbiati2025}. Spectroscopic data also enable more precise source classification, reducing selection biases in subsequent studies.

On the other hand, the rest-frame mid-IR continuum provides the most unbiased selection of galaxies and AGN. While hard X-rays (2--10\,keV or higher) might seem ideal for AGN selection, Compton-thick absorption ($N_\text{H} \gtrsim 10^{24}\,\text{cm}^{-2}$) significantly affects detection rates\cite{Bassani1999,Daddi2007,Treister2009}. Pioneer IRAS\cite{Spinoglio1995} and ISO mid-IR surveys\cite{Elbaz2002} demonstrated that mid-IR selection ($8 \lesssim \lambda \lesssim 15\,\mu$m) closely approximates bolometric flux-limited selection, recently confirmed for local AGN in the 12$\,\mu$m sample\cite{Spinoglio2024}. Therefore, FIRESS's mid-IR spectroscopic selection from a blind spectroscopic survey will provide an unbiased sample proportional to the bolometric power of both AGN and galaxies across $0 \lesssim z \lesssim 4$.

\begin{figure}
\centering
\includegraphics[width=0.8\textwidth]{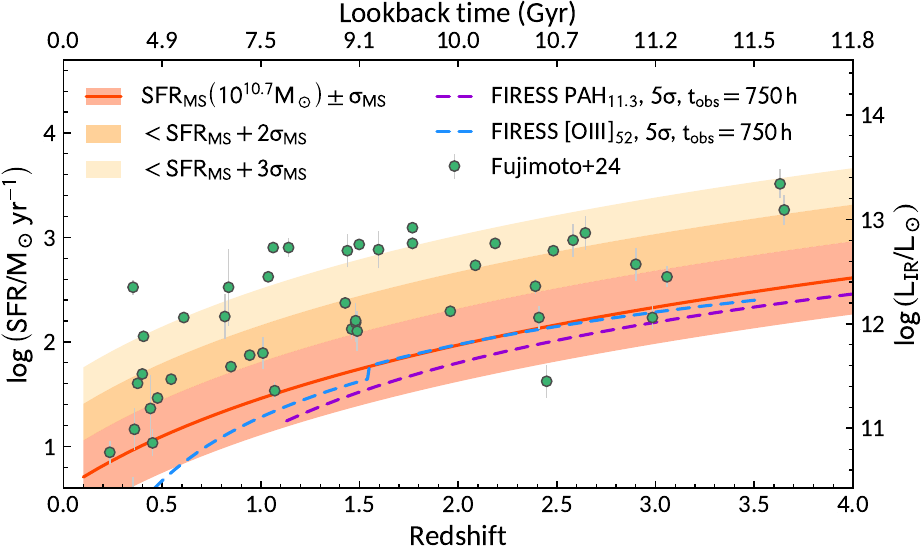}
\caption{Star Formation Rate (left y-axis, SFR in $\rm{M_\odot\,yr^{-1}}$) and total IR luminosity (right y-axis) as a function of redshift for a $10^{10.7}\, \rm{M_\odot}$ galaxy in the Main Sequence (MS, red solid line\cite{Scoville2017}). The red-shaded area shows the 0.35 dex intrinsic scatter around the MS\cite{Schreiber2015}, while the dark and light orange-shaded areas indicate the location of galaxies whose SFR is $+2\sigma$ and $+3\sigma$ above the MS, respectively. The blue dotted line shows the SFR (left y-axis) and infrared luminosity (right y-axis) of a low-metallicity galaxy detectable through the [\ion{O}{iii}]$_{52}$ line at $5\sigma$ for a 200\,arcmin$^2$ survey with 640\,h of exposure time with FIRESS low-resolution full-band mapping mode. The purple long-dashed line shows the same with the $11.3\, \rm{\mu m}$ PAH emission feature for a dusty star-forming galaxy. The green circles indicate the magnification-corrected infrared luminosities of galaxies detected in the ALCS survey\cite{Fujimoto2024} with a measured spectroscopic redshift, which show that most galaxies detected in this survey can be spectroscopically detected by FIRESS.}\label{fig_z_SFR}
\end{figure}

\begin{table}[b!]
\caption{A total number of $\sim 1000$ galaxies are expected to be detected at $\ge 5\sigma$ in the $11.3\, \rm{\mu m}$ PAH band and/or the [\ion{O}{iii}]$_{52}$ line, in a 200\,arcmin$^2$ survey with a total integration time of 640\,h. For this estimate we adopted the luminosity function derived from the ALCS survey\cite{Fujimoto2024}.}\label{tab_survey}
\vspace{0.2cm}
\centering
\begin{tabular}{lccccc}
  \bf $\log(L/\rm{L_\odot})$ & $0.1 < z < 1$ & $1 < z < 2$ & $2 < z < 3$ & $3 < z < 4$ & $z > 4$ \\
  \hline \\[-0.3cm]
  $>12.5$  &  0   &   2 &  7 &  6 & 5 \\ 
  12--12.5 &  7   &  53 & 66 & 20 & 2 \\
  11.5--12 &  45  & 252 & 63 &  0 & 0 \\
  $<11.5$  &  384 & 177 &  0 &  0 & 0 \\
  \hline
\end{tabular}
\end{table}

\begin{figure}
\centering
\includegraphics[width=0.8\textwidth]{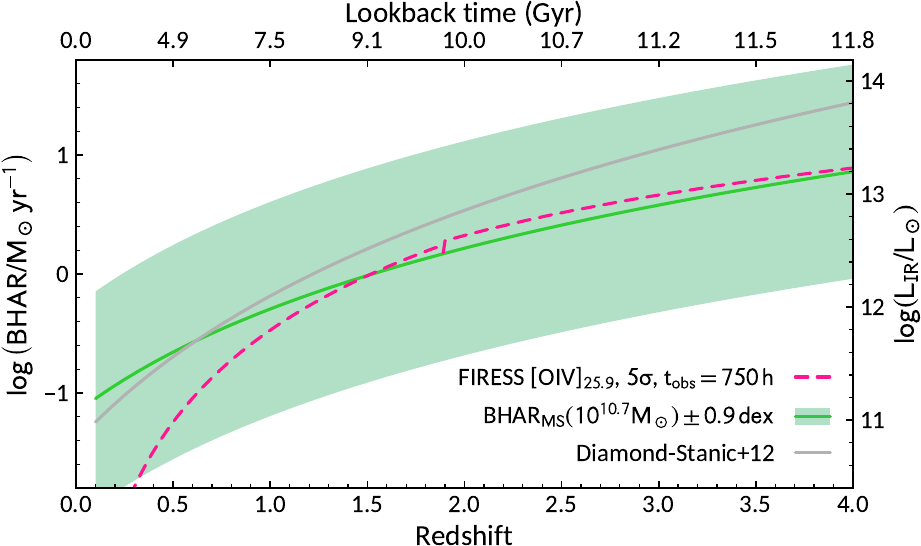}
\caption{Black hole accretion rate (BHAR; left axis) and total IR luminosity (right axis) versus redshift for a $10^{10.7}\,\text{M}_\odot$ main-sequence galaxy. The green line and shaded area show the BHAR derived from a SFR-AGN luminosity correlation obtained for X-ray selected AGN, at redshift z = 1.18-1.68, using SED decomposition, with 0.9\,dex scatter\cite{Setoguchi2021}. The grey solid line shows the estimate based on the BHAR-SFR relation obtained from [\ion{O}{iv}]$_{25.9}$ Spitzer observations for local Seyfert galaxies in the revised Shapley-Ames catalog\cite{Diamond-Stanic2012}. The pink dashed line indicates the $5\sigma$ detection limit for FIRESS in low-resolution full-band mapping mode for a sensitivity of $3.5 \times 10^{-19}\, \rm{W\,m^{-2}}$ ($5 \sigma$ at $24\, \rm{\mu m}$), using [\ion{O}{iv}]$_{25.9}$. FIRESS will measure BHAR through [\ion{O}{iv}]$_{25.9}$ emission up to $z = 4$.}\label{fig_z_BHAR}
\end{figure}

\section{Predictions of detections of SFR and BHAR}

To assess FIRESS's capabilities for studying galaxy evolution, we compare its sensitivity limits with predicted SFR and BHAR --\,the two key parameters driving galaxy evolution\,-- across cosmic time. Fig.\,\ref{fig_z_SFR} shows the SFR (left axis) as a function of redshift for a galaxy in the so called {\it  main sequence}\cite{Rodighiero2011,Elbaz2011}. We have assumed the stellar mass of a Milky Way-like galaxy ($10^{10.7}\, \rm{M_\odot}$), which is close to the knee of the galaxy mass function\cite{Baldry2012,Weigel2016} and thus represents the bulk of the galaxy population. The associated luminosity values (right axis) were adopted from the SFR--$L_\text{IR}$ relation, taking into account the evolution of the main sequence\cite{Scoville2017}. Line-luminosity calibrations for galaxies in the local Universe have shown that the brightest SFR tracers are the PAH emission band at 11.3\,$\mu$m, for dusty star-forming galaxies (DSFGs), and the [\ion{O}{iii}]$_{52}$ fine structure line for low-metallicity emission-line galaxies\cite{Mordini2021}. For the case assessment we also refer to the preparation of the Space Infrared telescope for Cosmology and Astrophysics (SPICA) \cite{Swinyard2009,Roelfsema2018} galaxy evolution science program \cite{Spinoglio2017,Spinoglio2021}.

Adopting a survey field size of $200\, \rm{arcmin^2}$, and assuming the current estimated sensitivity of FIRESS in low-spectral resolution mapping mode ($R > 85$) as provided by the PRIMA Exposure Time Calculator\footnote{\url{https://prima.ipac.caltech.edu/page/etc-calc}} (ETC), we expect to reach a sensitivity of $3.5 \times 10^{-19}\, \rm{W\,m^{-2}}$ ($5\sigma$) in 640\,h of exposure time, enabling the detection of approximately $\sim 1000$ galaxies, the majority of them during cosmic noon (see Table\,\ref{tab_survey}). Given the $\sim 3.3\, \rm{arcmin}$ separation on the sky between the slits of channels 1/3 and 2/4, a total coverage of $233\, \rm{arcmin^2}$ would be required to map a $23.3 \times 10\, \rm{arcmin^2}$ field, ensuring a common area of $20 \times 10\, \rm{arcmin^2}$ covered by all four channels. This configuration requires a total observing time of $750$\,h. The corresponding sensitivity at $24\, \rm{\mu m}$ is $3.5 \times 10^{-19}\, \rm{W\,m^{-2}}$ ($5\sigma$), assuming a spatial binning factor of $\sim 1.7$, which matches the pixel scale of the bluest channels ($7\farcs6$/px) to that of the reddest channel ($22\farcs9$/px) to ensure an homogeneous spatial resolution across the full spectral range.

The spectral resolution of $\sim$100 is adequate to detect not only the wide PAH feature at 11.3\,$\mu$m, but also the bright [\ion{O}{iii}]$_{52}$ line, as demonstrated by {\it Spitzer}-IRS observations of mid-IR emission lines using the low resolution mode\cite{Armus2004}. A total number of $\sim 600$--$900$ galaxies would be detectable in this survey through the PAH band at 11.3\,$\mu$m and/or the [\ion{O}{iii}]$_{52}$ emission line. The expected range corresponds to two extreme scenarios, assuming that all galaxies are dusty and bright PAH emitters (600), or low-metallicity and thus bright [\ion{O}{iii}]$_{52}$ emitters (900). The correlations adopted to estimate the mid-IR feature and line luminosities from the total IR luminosity have been taken from previous studies in the literature\cite{Mordini2021}.

Figure\,\ref{fig_z_SFR} shows magnification-corrected luminosities of galaxies with spectroscopic redshifts from the ALCS survey\cite{Fujimoto2024}. The proposed FIRESS blind-spectroscopic survey would provide mid- to far-IR rest-frame spectra for most of these galaxies up to $z \lesssim 4$. Recent JWST observations further demonstrate FIRESS's capabilities: studied ALMA-detected DSFGs at cosmic noon ($z \sim 2.3$--$2.7$) using NIRSPEC, finding infrared-luminous ($L_\text{IR} > 10^{12.4}\,\text{L}_\odot$), massive ($M_* > 10^{11}\,\text{M}_\odot$), and highly dust-obscured ($A_\text{V} \sim 3$--$4\,\text{mag}$) galaxies\cite{Cooper2025}. Similarly, MIRI observations of three DSFGs and a Lyman-break galaxy in an overdensity at $z = 4.05$ revealed large star formation rates ($300$--$2500\,\text{M}_\odot\,\text{yr}^{-1}$), substantial stellar masses ($M_* \sim 0.2$--$1.8 \times 10^{11}\,\text{M}_\odot$), and high extinction\cite{Crespo-gomez2024} ($A_\text{V} \sim 0.8$--$1.5\,\text{mag}$). Beyond cosmic noon, at $z \sim 4.8$, the DSFG XS55 was detected with SCUBA, ALMA, JWST, and tentatively with Chandra and XMM, revealing a massive main-sequence galaxy\cite{Sillassen2025} ($M_* \sim 5 \times 10^{10}\,\text{M}_\odot$, SFR $\sim 540\,\text{M}_\odot\,\text{yr}^{-1}$). All these galaxies fall within FIRESS detection limits. Thus, Fig.\,\ref{fig_z_SFR} demonstrates that FIRESS will be able to probe the redshift evolution of the SFR for average Milky Way-like galaxies in the Universe, the study of which is essential to understand galaxy evolution.

Figure\,\ref{fig_z_BHAR} shows the evolution of BHAR with redshift. We estimate this evolution using two approaches based on the known SFR evolution (Fig.\,\ref{fig_z_SFR}), assuming that the parallel evolution of SFR and BHAR, observed on a global scale\cite{madau_dickinson2014}, applies also to individual main-sequence galaxies. First, using the SFR-AGN luminosity correlation derived from X-ray selected galaxies in the redshift range z = 1.18-1.68, through SED decomposition fitting\cite{Setoguchi2021}, ($L_\text{AGN}/\text{erg}\,\text{s}^{-1}) \sim 10^{44} \times (\text{SFR}/\text{M}_\odot\,\text{yr}^{-1})$ with 0.9\,dex scatter (green line and shaded area in Fig.\,\ref{fig_z_BHAR}). Second, deriving the BHAR for a $10^{10.7}\, \text{M}_\odot$ main-sequence galaxy using the SFR-BHAR correlation from Spitzer [\ion{O}{iv}]$_{25.9}$ observations of Seyfert galaxies in the local Universe\cite{Diamond-Stanic2012} (gray line in Fig.\,\ref{fig_z_BHAR}).

Using a established correlation between the BHAR and the [\ion{O}{iv}]$_{25.9}$ emission-line luminosity\cite{Mordini2021}, we estimate FIRESS flux detection limits for a $3.5 \times 10^{-19}\, \rm{W\,m^{-2}}$, $5\sigma$ sensitivity (pink-dashed line). The proposed blind spectroscopic survey with FIRESS will detect AGN above the estimated knee of the luminosity function ($L^*_\text{IR}$) up to $z \sim 4$ using [\ion{O}{iv}]$_{25.9}$, as shown by Fig.\,\ref{fig_z_BHAR} for optimistic\cite{Diamond-Stanic2012} and pessimistic\cite{Setoguchi2021} predictions. Note that these BHARs represent instantaneous accretion rates during the AGN phase of these galaxies, not cosmic time-averaged values. Regarding the estimated number of source detections, if the parallel evolution of star formation and black hole accretion across cosmic time originates from the interplay between these two phenomena, this suggests that the number density of AGN and AGN-star formation composite sources is expected to be similar --\,within an order of magnitude\,-- to the observed number density of star forming galaxies\cite{Fujimoto2024}. Thus, we expect a considerable fraction of sources in Table\,\ref{tab_survey} ($\sim 20$--$50$\%) to host an active nucleus and be detectable through the [\ion{O}{iv}]$_{25.9}$ line.

\section{FIRESS Deep Follow-up observations: Chemical Evolution and AGN Feedback}

\begin{figure}
\centering
\includegraphics[width=\textwidth]{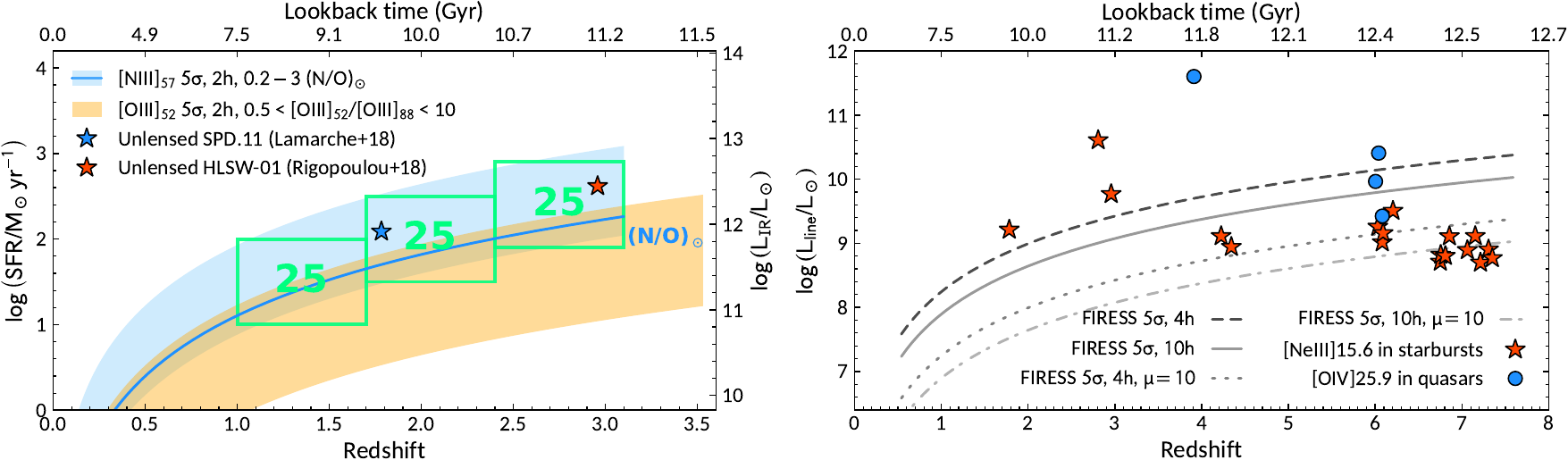}
\caption{\textit{Left:} FIRESS detection limits ($5\sigma$, 2\,h, low-res.) for [\ion{N}{iii}]$_{57}$ and [\ion{O}{iii}]$_{52}$ on the SFR-redshift plane. Blue-shaded area shows [\ion{N}{iii}]$_{57}$ sensitivity based on the $L_\text{IR}$--[\ion{O}{iii}]$_{88}$ correlation\cite{Mordini2021} for N/O = 0.2--3\,$\text{(N/O)}_\odot$\cite{Spinoglio2022a}, with the blue line indicating solar N/O = -0.86\,dex\cite{Asplund2021}. Orange-shaded area shows [\ion{O}{iii}]$_{52}$ sensitivity for density-dependent [\ion{O}{iii}]$_{52}$/[\ion{O}{iii}]$_{88}$ ratios (0.5--10)\cite{jafo2016,Spinoglio2022a}. Stars show magnification-corrected luminosities of lensed galaxies with [\ion{N}{iii}]$_{57}$ detections\cite{Lamarche2018,Rigopoulou2018}. Green boxes indicate the proposed observational program coverage. \textit{Right:} Expected [\ion{Ne}{iii}]$_{15.6}$ fluxes in star-forming galaxies (red stars) and [\ion{O}{iv}]$_{25.9}$ in quasars (blue circles) based on high-z [\ion{O}{iii}]$_{88}$ detections\cite{Ferkinhoff2010,Lamarche2018,Rigopoulou2018,Walter2018,DeBreuck2019,Hashimoto2019a,Hashimoto2019b,Harikane2020,Witstok2022,Ren2023,Algera2024}, assuming typical line ratios from local galaxies\cite{Cormier2015,Cormier2019,jafo2016}. Gray lines show FIRESS sensitivity limits for different integration times and magnifications ($5\sigma$: 4\,h dashed, 10\,h solid, 4\,h with $\mu = 10$ dotted, 10\,h with $\mu = 10$ dot-dashed).}\label{fig_z_SFR_N3O3}
\end{figure}

High-quality spectra from FIRESS follow-up observations of selected sources will address fundamental questions in galaxy evolution: the build-up of metals and feedback from active SMBHs through molecular and ionized gas outflows. While metallicity studies can use FIRESS's low-resolution spectroscopy, feedback analyses require the high-resolution mode ($R > 2000$). Targets can be drawn from the proposed 200\,arcmin$^2$ blind spectroscopic survey, PRIMAger deep imaging surveys\cite{Burgarella2024}, or catalogs of high-z sources from JWST, ALMA, and future facilities.

The chemical nature of DSFGs remains controversial even at low redshift\cite{Herrera-Camus2018,Chartab2022,Perez-Diaz2024}. Independent O/H and N/O measurements in galaxies can provide crucial constraints on chemical enrichment processes, potentially revealing past metal-poor gas accretion events\cite{Amorin2010,Spinoglio2022a,Perez-Diaz2024}. N/O estimates based on optical [\ion{N}{ii}]$_{\rm \lambda \lambda 6548,6583}$ or far-IR [\ion{N}{ii}]$_{122,205}$ lines can be contaminated by low-ionization diffuse gas that may not reflect the composition of ongoing star forming regions\cite{Peng2021}. While [\ion{N}{iii}]$_{57}$ observations are essential for robust N/O ratios, fewer than 50 local galaxies have Herschel or SOFIA measurements, and neither JWST nor ALMA can access this line at cosmic noon. At these redshifts, [\ion{N}{iii}]$_{57}$ detections exist for only two gravitationally-lensed galaxies: SDP.11 at $z = 1.78$\cite{Lamarche2018} and HLSW-01 at $z = 2.96$\cite{Rigopoulou2018}. For unlensed galaxies, even stacked Herschel data yields only upper limits for N/O ratios\cite{Wardlow2017}.

FIRESS low-resolution mode ($R > 85$) will detect [\ion{N}{iii}]$_{57}$ and [\ion{O}{iii}]$_{52}$ lines in main-sequence galaxies from the local Universe to cosmic noon ($z \le 3.1$). Fig.\,\ref{fig_z_SFR_N3O3} (left panel) shows the expected FIRESS sensitivity ($5\sigma$, 2\,h, low-res. full-band spectrum) for [\ion{N}{iii}]$_{57}$ emission (blue-shaded area), based on the $\log L_\text{IR}$--[\ion{O}{iii}]$_{88}$ correlation for low-metallicity galaxies\cite{Mordini2021} and N/O abundances of $0.2$--$3\,\text{(N/O)}_\odot$\cite{Spinoglio2021}. Blue and red stars indicate magnification-corrected luminosities for the two lensed galaxies with [\ion{N}{iii}]$_{57}$ detections\cite{Lamarche2018,Rigopoulou2018}. A proposed observing program could target 75 DSFGs across three redshift bins: 25 galaxies with $L_\text{IR} > 10^{10.8}\,\text{L}_\odot$ at $1 < z < 1.6$, 25 with $L_\text{IR} > 10^{11.3}\,\text{L}_\odot$ at $1.6 < z < 2.4$, and 25 with $L_\text{IR} > 10^{11.7}\,\text{L}_\odot$ at $2.4 < z < 3.1$. Assuming a sensitivity of $1.9 \times 10^{-19}\,\text{W}\,\text{m}^{-2}$ ($5\sigma$, 1\,h per spectral setting; PRIMA ETC), the [\ion{N}{iii}]$_{57}$ line can be detected in main-sequence cosmic noon galaxies with $\sim 1$--$2$\,h integration per target. The total program needs $\sim 200$\,h plus 20\% overheads, with two exposures per target covering the full spectral range.

\begin{figure}
\centering
\includegraphics[width=\textwidth]{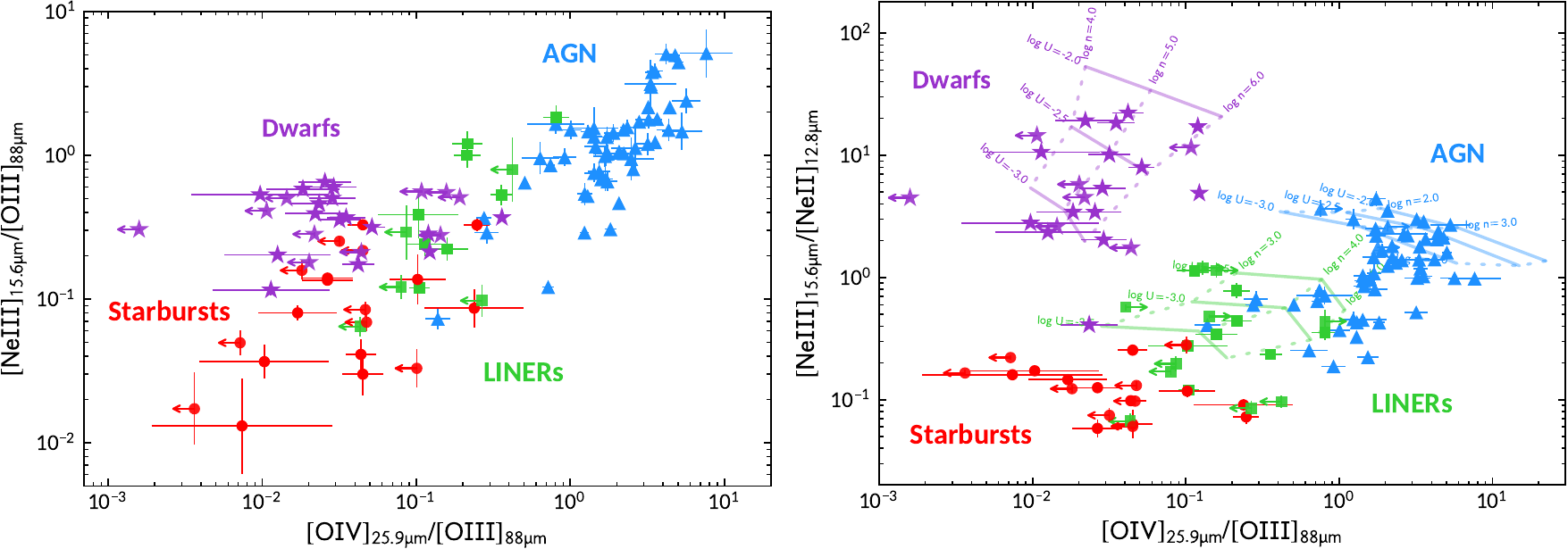}
\caption{Predictions for the [\ion{Ne}{iii}]$_{15.6}$ and [\ion{O}{iv}]$_{25.9}$ line intensities at high-z are based on the observed ratios, relative to the [\ion{O}{iii}]$_{88}$ (left panel) and [\ion{Ne}{ii}]$_{12.8}$ lines (right panel), measured by Spitzer and Herschel for different types of galaxies in the local Universe, including low-metallicity star-forming dwarf galaxies\cite{Cormier2015,Cormier2019} (violet stars), starburst galaxies (red stars), LINERs (green squares) and AGN\cite{jafo2016,Spinoglio2017} (blue triangles). The dotted and solid lines on the right panel represent the photoionization predictions by CLOUDY\cite{Ferland2013} for low-metallicity star-forming galaxies (violet), low-luminosity AGN (green) and Seyfert nuclei (blue).}\label{fig_lowzratios}
\end{figure}

Beyond N/O ratios, O/H abundances can be obtained using FIRESS's suite of nebular IR lines. Local Universe observations demonstrate that robust metallicities for both star-forming galaxies and AGN can be derived using the brightest IR lines\cite{jafo2021,Perez-Diaz2022}. While N/O measurements extend to cosmic noon ($z \le 3.1$), O/H abundances can be determined for luminous or gravitationally-lensed galaxies up to the reionization epoch. Fig.\,\ref{fig_z_SFR_N3O3} (right panel) shows expected fluxes for [\ion{Ne}{iii}]$_{15.6}$ in star-forming galaxies (red stars) and [\ion{O}{iv}]$_{25.9}$ in quasars (blue circles), based on high-z [\ion{O}{iii}]$_{88}$ detections\cite{Ferkinhoff2010,Lamarche2018,Rigopoulou2018,Walter2018,DeBreuck2019,Hashimoto2019a,Hashimoto2019b,Harikane2020,Witstok2022,Ren2023,Algera2024} and typical line ratios ([\ion{Ne}{iii}]$_{15.6}$/[\ion{O}{iii}]$_{88}$ $\sim 0.4$ and [\ion{O}{iv}]$_{25.9}$/[\ion{O}{iii}]$_{88}$ $\sim 4$) observed in local low-metallicity star-forming\cite{Cormier2015,Cormier2019} and Seyfert galaxies\cite{jafo2016} (Fig.\,\ref{fig_lowzratios}). Metallicity estimates can be obtained with, e.g. HII-CHI-mistry-IR \cite{Perez-Montero2014,jafo2021,Perez-Diaz2022}, which analyzes multiple emission lines ([\ion{Ne}{ii}]$_{12.8}$, [\ion{Ne}{iii}]$_{15.6}$, [\ion{S}{iii}]$_{18.7,33.5}$, [\ion{S}{iv}]$_{10.5}$, [\ion{O}{iii}]$_{52,88}$, and [\ion{N}{iii}]$_{57}$) while accounting for ionization parameter and N/O ratios. Though some lines fall outside FIRESS coverage at specific redshifts (e.g., [\ion{S}{iv}]$_{10.5}$ at $z < 1.3$, [\ion{O}{iii}]$_{88}$ at $z > 1.6$, [\ion{O}{iii}]$_{52}$ and [\ion{N}{iii}]$_{57}$ at $z > 3.1$), the code flexibly uses available lines for best-effort metallicity determinations.

FIRESS will also play a pivotal role in studying the feedback mechanisms that regulate galaxy evolution. Feedback episodes often manifest as multiphase gas outflows, with the molecular component carrying most of the mass, directly affecting the star-forming reservoir. These outflows can be identified in high-resolution spectra through P-Cygni profiles and absorption blueshifted wings in molecular lines such as the OH doublets at $119$ and $79\, \rm{\mu m}$\cite{Gonzalez-Alfonso2012,Spoon2013,Gonzalez-Alfonso2017b}, and through emission line wings of highly ionized emission lines such as [\ion{O}{iv}]$_{25.9}$ and [\ion{Ne}{v}]$14.3,24.3\, \rm{\mu m}$. These measurements will provide the mass, energetics, and kinematics of outflows across different galaxy evolutionary stages. By observing a sample of galaxies with high star-formation rates and AGN activity, FIRESS will disentangle the roles of starburst-driven winds and AGN-driven outflows in quenching star formation. Combining these results with observations from facilities like JWST and ALMA will enable a comprehensive multiphase characterization of feedback processes, revealing how galaxies transition from gas-rich, star-forming systems to passive, red-and-dead states.

\section{Beyond the cosmic noon}

The unprecedented sensitivity of FIRESS opens new frontiers for exploring the early Universe through rest-frame mid-IR spectroscopy. As shown in Fig.\,\ref{fig_z_SFR_N3O3} (right panel), bright unlensed galaxies and moderately lensed sources ($\mu \gtrsim 10$) like those in the ALCS survey\cite{Fujimoto2024} are within reach with just a few hours of integration time. 

FIRESS observations of galaxies at $z > 4$ would be particularly powerful for understanding the drivers of cosmic reionization and early chemical enrichment. For extreme Ly$\alpha$ emitters, the [\ion{Ne}{iii}]$_{15.6}$ line combined with other high-ionization transitions like [\ion{O}{iv}]$_{25.9}$ would provide crucial constraints on the ionizing continuum shape and escape fraction of ionizing photons. In sources showing anomalously high N/O ratios and extreme densities at $z > 4$, similar to recent JWST discoveries at $z > 10$\cite{Topping2024,Topping2025}, rest-frame mid-IR metallicity diagnostics would be invaluable for probing chemical enrichment processes in high-density environments where optical tracers may be compromised. For instance, pointed observations of a small sample of $\sim 20$ high-z targets, selected among bright and lensed systems, with $\sim 2$\,h of exposure per spectral setting, would provide the first restframe mid-IR spectra of these galaxies in a total of $80$\,h (Fig.\,\ref{fig_z_SFR_N3O3}, right).

Furthermore, the recent discovery of Little Red Dots\cite{Matthee2024} (LRDs) has revealed a potentially large population of X-ray and radio elusive AGN at high redshift\cite{Yue2024,Perger2025}, which are also difficult to identify using classical diagnostics based on optical line ratios\cite{Mazzolari2024}. These sources are thought to host massive black holes\cite{Durodola2025,Graham2025} ($\sim 10^6$--$10^8\, \rm{M_\odot}$) accreting close to their Eddington limit\cite{Inayoshi2025}, implying a BHAR distribution in the range $\log(\text{BHAR}/\rm{M_\odot\,yr^{-1}}) \sim -1.6$--$0.4$. Assuming that LRDs follow local correlations between BHAR and high-ionization line fluxes, such as the one adopted in this work, [\ion{O}{iv}]$_{25.9}$ emission should be detectable in the brightest LRDs at cosmic noon, while detections at higher redshifts will likely require gravitational lensing. For instance, a magnification factor of $\mu = 10$ would enable FIRESS to detect LRDs at redshifts $z > 4$ (Fig.\,\ref{fig_z_BHAR}). The detection of high-ionization lines in LRDs accessible to FIRESS --\,including local analogues\cite{Bisigello2025}\,-- will be essential for characterizing the nature, BHAR, and chemical properties of these enigmatic sources, providing new insights into black hole growth and AGN phenomena in the early Universe.

In this context, emission line ratio diagrams of galaxies beyond cosmic noon, such as those presented in Fig.\,\ref{fig_lowzratios} for local galaxies, can help identify the main gas excitation mechanisms, whether driven by AGN or star formation processes, and discriminate between solar and low-metallicity environments. For instance, by measuring the [\ion{O}{iv}]$_{25.9}$ and [\ion{Ne}{iii}]$_{15.6}$ lines with FIRESS for galaxies with available or upcoming [\ion{O}{iii}]$_{88}$ ALMA detections\cite{Ferkinhoff2010,Lamarche2018,Rigopoulou2018,Walter2018,DeBreuck2019,Hashimoto2019a,Hashimoto2019b,Harikane2020,Witstok2022,Ren2023,Algera2024}, we can classify galaxies using IR emission-line diagnostics (Figure,\ref{fig_lowzratios}).

\section{Summary}

PRIMA, equipped with the FIRESS spectrometer, represents a transformative step forward in the study of galaxy evolution. By offering unprecedented sensitivity and broad wavelength coverage in the mid- and far-IR, PRIMA will enable simultaneous measurements of star formation rates, black hole accretion rates, and chemical abundances in galaxies. A blind spectroscopic survey similar to that proposed in this study (200\,arcmin$^2$ in a total of 750\,h observing time) will provide star-formation rates --\,through the PAH 11.3\,$\mu$m band and the [\ion{Ne}{iii}]$_{15.6}$ emission-line\,-- and black hole accretion rates --\,through the [\ion{O}{iv}]$_{25.9}$ line\,-- for hundreds of galaxies. This will deliver a complete and unbiased census of heavily obscured star-forming galaxies and AGN from the local Universe to beyond cosmic noon, filling the critical gap between JWST and ALMA.

FIRESS's capability to observe key IR fine-structure lines will allow detailed investigations of AGN feedback, metal enrichment, and the processes driving galaxy evolution through follow-up spectroscopic observations. N/O ratios and N/O-independent O/H abundances will provide a unique view of the chemical evolution in galaxies, probing gas-accretion and metal removal processes. High-resolution spectroscopy will reveal massive molecular outflows and highly ionized winds, testing star-formation quenching scenarios. These measurements will reveal how star formation and black hole growth peaked during cosmic noon and subsequently declined, while also shedding light on the interplay between these processes and their impact on the ISM.

Beyond cosmic noon, PRIMA will pioneer mid-IR spectroscopy of galaxies in the reionization epoch, unveiling the properties of the primary ionizing continuum in bright Lyman-$\alpha$ emitters and low-metallicity AGN. For instance, FIRESS can detect the [\ion{Ne}{iii}]$_{15.6}$ line in star-forming galaxies (SFR\,$\sim 500\, \rm{M_\odot \,yr^{-1}}$) or the [\ion{O}{iv}]$_{25.9}$ line in AGNs ($L_\text{bol} \sim 10^{13}\, \rm{L_\odot}$) at $z \sim 6$ using an integration time of 2 hours. Much fainter galaxies will be accessible through present and upcoming catalogs of lensed sources.

By leveraging its unique strengths, PRIMA/FIRESS will address fundamental questions about the nature of obscured galaxies and AGN, their role in shaping the evolution of the Universe, and the physical processes governing their transformation across cosmic time. This mission will set a new benchmark for the exploration of galaxy evolution in the IR.

\subsection*{Disclosures}
The authors declare that there are no financial interests, commercial affiliations, or other potential conflicts of interest that could have influenced the objectivity of this research or the writing of this paper.

\subsection*{Code and Data Availability}
All data in support of the findings of this paper are available within the article or as supplementary material.

\subsection*{Acknowledgments} 
The authors would like to thank the referees for their careful reading of the manuscript and for their constructive suggestions. JAFO acknowledges financial support by the Spanish Ministry of Science and Innovation (MCIN/ AEI/10.13039/501100011033), by ``ERDF A way of making Europe'' and by ``European Union NextGenerationEU/PRTR'' through the grants PID2021-124918NB-C44 and CNS2023-145339; MCIN and the European Union -- NextGenerationEU through the Recovery and Resilience Facility project ICTS-MRR-2021-03-CEFCA. This work made use of \textsc{Astropy} (\url{http://www.astropy.org}): a community-developed core Python package and an ecosystem of tools and resources for astronomy\cite{astropy22}.


\bibliography{firess}   
\bibliographystyle{spiejour}   





\end{document}